\pdfoutput=1
\documentclass{Interspeech}

 \interspeechcameraready

\title{Speech DF Arena: A Leaderboard for Speech DeepFake Detection Models}

\author{
  Sandipana Dowerah$^{1}$, Atharva Kulkarni$^{2}$, Ajinkya Kulkarni$^{3}$, Hoan My Tran$^{4}$, Joonas Kalda$^{1}$, Artem Fedorchenko$^{1}$, Benoit Fauve$^{6}$, Damien Lolive $^{5}$, Tanel  Alumäe$^{1}$, Matthew Magimai Doss$^{3}$ \\
  $^{1}$Tallinn University of Technology, Estonia,
  {$^{2}$ MBZUAI, UAE},
  {$^{3}$ Idiap Research Institute, Switzerland},
  {$^{4}$ Univ. Rennes, CNRS, Irisa, France},
  {$^{5}$ Univ. Bretagne Sud, CNRS, Irisa, France},
  {$^{6}$ Validsoft Ltd., UK}
  }

\email{firstname.lastname@taltech.ee$^{1}$, firstame.lastname@mbzuai.ac.ae$^{2}$, firstname.lastname@idiap.ch$^{3}$, firstname.lastname@irisa.fr$^{4,5}$,
firstname.lastname@validsoft.com$^{6}$}


\usepackage{comment}
\usepackage{comment}
\usepackage{times}
\usepackage{latexsym}
\usepackage{amsmath}
\usepackage[T1]{fontenc}
\usepackage{multicol}
\usepackage{hhline}
\usepackage[utf8]{inputenc}
\usepackage{microtype}
\usepackage{inconsolata}
\usepackage{graphicx}
\usepackage{url}
\usepackage{colortbl}
\usepackage{multirow}
\usepackage{lineno}
\usepackage{booktabs}
\usepackage{hyperref}
\usepackage{cleveref}
\usepackage{xcolor}
\usepackage{todonotes}

\usepackage[utf8]{inputenc}
\usepackage[T1]{fontenc}
\usepackage{graphicx}
\DeclareGraphicsExtensions{.pdf,.png,.jpg}

\makeatletter
\newcommand*\iftodonotes{\if@todonotes@disabled\expandafter\@secondoftwo\else\expandafter\@firstoftwo\fi}

\begin{document}

\maketitle


\begin{abstract}
    Parallel to the development of advanced deepfake audio generation, audio deepfake detection has also seen significant progress. However, a standardized and comprehensive benchmark is still missing. To address this, we introduce Speech DeepFake (DF) Arena, the first comprehensive benchmark for audio deepfake detection. Speech DF Arena provides a toolkit to uniformly evaluate detection systems, currently across $14$ diverse datasets and attack scenarios, standardized evaluation metrics and protocols for reproducibility and transparency. It also includes a leaderboard to compare and rank the systems to help researchers and developers enhance their reliability and robustness. We include $14$ evaluation sets, $12$ state-of-the-art open-source and $3$ proprietary detection systems. 
    Our study presents many systems exhibiting high EER in out-of-domain scenarios, highlighting the need for extensive cross-domain evaluation. The leaderboard is hosted on Huggingface\footnote{\url{https://huggingface.co/spaces/Speech-Arena-2025/Speech-DF-Arena}} and a toolkit for reproducing results across the listed datasets is available on GitHub\footnote{\url{https://github.com/Speech-Arena/speech_df_arena}}.  
\end{abstract}

\section{Introduction}

As AI-powered audio deepfakes continue to evolve, distinguishing between genuine and synthetic speech is becoming more challenging. The technological advancement achieved by the current text-to-speech (TTS) \cite{ss1,ss2,ss3, ElHajal_NAACL_2025} and voice conversion (VC) \cite{ssvc1,ssvc2} systems to obtain realism has also given rise to generating audio deepfakes. Deepfakes are imitations of a person's voice, image, or video using deep neural network (DNN) techniques. Deepfakes are being exploited for malicious and criminal activities; for instance, audio deepfakes have been used to imitate Italian defence minister Guido Crosetto in a bid to get tycoons to pay millions of ransoms for the release of Italian journalists overseas\footnote{https://www.theguardian.com/world/2025/feb/10/ai-phone-scam-targets-italian-business-leaders-including-giorgio-armani}. Similar methods have reportedly been used to scam the CEO of an energy-based company in the UK\footnote{https://www.forbes.com/sites/jessedamiani/2019/09/03/a-voice-deepfake-was-used-to-scam-a-ceo-out-of-243000/}, as well as the infamous Biden's robocalls used during the US elections \cite{kulkarni2024generalization}. Moreover, children's speech being recorded by smart devices knowingly or unknowingly is highly sensitive, remaining underexplored in anonymization \cite{cvp}, making them vulnerable to exploitation in audio deepfakes. This underscores the urgent need for robust audio deepfake detection systems capable of identifying and mitigating these threats. 

Although recent studies have demonstrated the increasing difficulty of audio deepfake detection \cite{Mai_2023, yi}, newly published work offers a promising solution \cite{kulkarni2024generalization, zhang2024towards,kulkarni25_interspeech, todisco2019asvspoof2019futurehorizons, liu2023asvspoof, wang2025asvspoof5, yi2022add}. Several audio deepfake detection challenges are introduced, such as the ASVspoof series \cite{tak21_asvspoof, wang2025asvspoof5,kinnunen17_interspeech,todisco2019asvspoof2019futurehorizons} and the Audio Deepfake Detection (ADD) challenge series \cite{yi2022add,yi2023add}, where participants are asked to submit scores for evaluation on a hidden test set. The effectiveness of these systems remains unproven for generalization in unseen scenarios. Despite these rapid advancements in detecting audio deepfakes, a unified, comprehensive benchmark for comparing various detectors is still lacking.

VoiceWukong \cite{yan2024voicewukong} generates a comprehensive voice deepfake dataset using commercial and open-source tools, benchmarking $12$ SOTA detection systems. \cite{zhang2024towards} benchmarks various continual learning frameworks on a Multilayer perceptron (MLP) model to evaluate performance on newly created deepfake attacks while maintaining performance on existing ones. CtrSVDD \cite{Zang2024CtrSVDDAB} focuses on singing voice detection by creating a comprehensive singing deepfake dataset, evaluating $14$ deepfake detection methods across various singing voices. Most benchmarks evaluate multiple models on a single dataset, limiting their ability to assess the generalizability of SOTA detection systems across diverse datasets. This gap between academic research and real-world scenarios is critical, as robustness against unseen out-of-distribution attacks is essential for practical applications. Also, existing benchmarks lack a unified evaluation platform for comparing speech deepfake detection systems across various datasets and attack types. Inconsistent metrics and protocols make performance comparisons difficult and unreliable.

To address this gap, we introduce Speech DF Arena, inspired by the success of TTS-arena\footnote{\url{https://huggingface.co/spaces/TTS-AGI/TTS-Arena}} and ASR-arena\footnote{\url{https://huggingface.co/spaces/hf-audio/open_asr_leaderboard}}, an innovative platform designed to compare and evaluate deepfake detection systems. One of the objectives behind Speech DF Arena is to address the challenge of accurately evaluating the performance of audio deepfake detection systems. Some metrics may not account for the complexity of new and emerging deepfake techniques, while others may rely on small datasets or limited test scenarios. We have included equal error rate (EER), pooled EER, accuracy and F1 score for evaluation. By incorporating community participation into the ranking process, the Speech DF Arena aims to address these limitations and provide a more comprehensive assessment of deepfake detection systems. In this phase, we include $14$ datasets, $12$ SOTA open-source and $3$ proprietary detection systems described in Section \ref{dataset}. Experimentation is described in Section \cref{experi}, results are in Section \ref{results}, and Section \ref{conclude} concludes with future direction.

\begin{table*}[!ht]
\centering
\caption{Summary of the open-source SOTA deepfake detectors. N/A indicates the non-availability of the repositories.}
\resizebox{\textwidth}{!}{%
\begin{tabular}{l|c|c|c|c|c}
\toprule
& \multicolumn{1}{c|}{\textbf{Detectors}}  &
\multicolumn{1}{c|}{\textbf{Feature Extractor}} & \multicolumn{1}{c|}{\textbf{Classifier}} & %\multicolumn{1}{c}{ASV21DF} & \multicolumn{1}{c}{ASV24-E} & 
\multicolumn{1}{c|}{\textbf{Repositories}} & \multicolumn{1}{c}{\textbf{References}} \\ 
%& \multicolumn{1}{c}{ADD23-R1} 

\midrule

\parbox[t]{2mm}{\multirow{6}{*}{\rotatebox[origin=c]
   {90}{\textbf{SSL}}}}
& XLS-R + Sensitive Layer Selection (XLSR SLS)& XLS-R & MLP & https://github.com/QiShanZhang/SLSforASVspoof-2021-DF & \cite{zhang2024audio} \\
& XLS-R + Conformer + Temporal-Channel Modeling (TCM) & XLS-R & Conformer & https://github.com/ductuantruong/tcm\_add & \cite{truong2024temporal} \\
 & XLS-R + Mamba& XLS-R& Mamba& https://github.com/swagshaw/XLSR-Mamba&\cite{xlsrMamba}\\
& WavLM-ECAPA-TDNN & WavLM & ECAPA-TDNN & N/A & \cite{kulkarni2024generalization}\\
& Whisper-DF-MesoNet-MFCC & Whisper + MFCC & MesoNet & https://github.com/piotrkawa/deepfake-whisper-features & \cite{kawa2023deepfake} \\
& Hubert-ECAPA-TDNN & HuBERT & ECAPA-TDNN & N/A & \cite{kulkarni2024generalization}\\
& Wav2Vec2-ECAPA-TDNN & Wav2Vec2 & ECAPA-TDNN & N/A & \cite{kulkarni2024generalization}\\
 & Wav2Vec2-AASIST& Wav2Vec2& AASIST& https://github.com/TakHemlata/SSL\_Anti-spoofing&\cite{tak2022automatic}\\
 & Nes2Net& Wav2Vec2&Res2Net & https://github.com/Liu-Tianchi/Nes2Net&\cite{liu2025nes2net}\\
\midrule
\parbox[t]{2mm}{\multirow{2}{*}{\rotatebox[origin=c]
   {90}{\textbf{GNN}}}}
& AASIST-Large& Waveform & Spectro-Temporal Graph Attention Networks & https://github.com/clovaai/aasist & \cite{jung2022aasist}  \\
& RawGAT-ST Multiplicative Fusion& Waveform & Spectro-Temporal Graph Attention Networks & https://github.com/eurecom-asp/RawGAT-ST-antispoofing & \cite{tak21_asvspoof} \\
\midrule
\parbox[t]{2mm}{\multirow{2}{*}{\rotatebox[origin=c]
   {90}{\textbf{CNN}}}}& RawNet-2 & Sinc filters & Gated Recurrent Units & https://github.com/eurecom-asp/rawnet2-antispoofing & \cite{tak2021end} \\
 & & & & &\\
%\midrule
%  \parbox[t]{2mm}{\multirow{3}{*}{%  {90} {Prop.}}} & Syntra &  &  &  & \\
 %  & Resemble detect & & & & \\ 
 %  & Whispeak &  &  &  & \\
\bottomrule
\end{tabular}%
}
\label{tab:detectors}
\end{table*}

% TODO: We have created benchmrk over datasets and models as well, exisiting work only foocus on benchmark on models

\section{Speech DF Arena: a unified benchmark}
\label{dataset}

% \begin{itemize}
%     \item Justification for a standardized, evolving benchmark
%     \item Discussion on dataset diversity and realistic evaluation
%     \item Description of the Hugging Face leaderboard/Details on datasets and SOTA models included
%     \item /Methodology for evaluation and scoring/Future plans for expansion and improvements
%     \item Possible analysis of current datasets
% \end{itemize}

% \begin{figure}[htb!]
%     \centering
%     \includegraphics[width=\columnwidth]{Speech_DF_Arena.png}
%     \caption{Framework}
%     \label{fig:framework}
% \end{figure}

% on which models all models are trained, >>> TODO

\subsection{Datasets}
\label{sets}

%\subsubsection{\textbf{\textit{Datasets}}}
Speech DF Arena in Phase I comprises $14$ different widely recognized and extensively used evaluation datasets that encompass common deepfake attacks. We use the ASVspoof series (2019, 2021 and 2024)\cite{todisco2019asvspoof2019futurehorizons, liu2023asvspoof, Wang2024_ASVspoof5}. ASVspoof 2019 is primarily built for speaker verification and includes generated speech. ASVspoof 2021 builds upon 2019 to introduce a special deepfake track. ASVspoof 2024 further adds more modern attacks. We also use the evaluation sets from the ADD challenges (2022 and 2023; Track 1 and 3; Round 1 and 2) \cite{yi2022add, yi2023add}, which consist of English and Chinese deepfake audios. For neural audio codec and vocoder-based attacks, we use CodecFake \cite{xie2024codecfake} and LibriSeVoc \cite{sun2023ai} datasets. SONAR \cite{li2024sonarsyntheticaiaudiodetection} and Fake or Real (FoR) \cite{reimao2019dataset} datasets offer deepfake speech from some of the most recent end-to-end TTS models. DFADD \cite{du2024dfadd} included attacks generated using recent advancements in diffusion and flow-matching-based TTS models. The In-the-wild dataset \cite{muller2022does} includes real-world deepfake voices for celebrities and other public figures mined from social media and video streaming platforms.

\subsection{{{Detection systems}}}

We categorised the SOTA detection systems as SSL-based, Graph Neural Network (GNN)-based, and Convolutional Neural Network (CNN)-based systems as detailed in \cref{tab:detectors}. We selected the systems based on two criteria. First, SOTA performances on SSL-based approaches, published in top-tier conferences, and challenges organised for audio deepfake detection, such as ADD and ASVspoof. Secondly, we selected the models with the minimum pre-processing steps and SOTA end-to-end approaches where the system takes input directly as a raw waveform. SSL-based models utilise a transformer architecture and representations learned from large speech foundation models as their backbone, while adding a front-end classification network on top. GNN-based models are relatively smaller in size and use Graph Attention to identify spectral and temporal cues for detection \cite{tak21_asvspoof}. The SSL-based models leverage large amounts of pre-training data to encode relevant features \cite{kulkarni2024generalization}, and GNN-based models are lightweight and suitable for real-world applications. CNN-based models utilize convolutional layers to directly process raw waveforms, capturing spectral and temporal features without requiring hand-crafted feature extraction. 

The selected systems are trained on a wide range of attacks generated using TTS, VC, neural audio codecs, vocoders and In-the-wild scenarios, which were chosen for their proven effectiveness and innovation in tackling the challenges of detecting audio deepfakes. The platform includes a leaderboard that displays the highest-ranked detection systems. We have also developed a DF Arena toolkit, available on GitHub. The toolkit offers a unified interface for computing evaluation metrics, namely EER, pooled EER, F1 score and accuracy, across open-source models on any dataset by adapting to a standardized protocol format.

The XLSR-Mamba \cite{xlsrMamba} system submitted by the authors to the DF arena leverages cross-lingual self-supervised speech representations from the XLSR architecture. By capturing rich acoustic and phonetic features across multiple languages and speakers, XLSR-Mamba can detect subtle artefacts introduced by both TTS and VC systems. XLSR-Mamba emphasizes robust feature extraction and generalization, enabling strong performance on curated datasets while maintaining reasonable resilience to real-world variability, including partial manipulations and codec distortions. XLSR-Mamba thus represents a hybrid approach, combining the power of pretrained embeddings with task-specific fine-tuning for effective deepfake detection.

Three proprietary systems have been submitted to the DF Arena: Whispeak, Syntra.io, and Resemble AI. Each detection system developed by industry providers specializing in speech security and synthetic media detection contributed to the benchmark. Their inclusion is significant, as they represent commercial-grade approaches that are actively deployed in real-world applications, complementing the open-source and academic baselines. By participating in the Arena, proprietary systems provide valuable reference points for the state of industrial practice, allowing the research community to assess how commercial detectors compare with publicly available SOTA systems under standardized evaluation protocols.

%We select $9$ such popular open-source deepfake detection systems for benchmarking covering different model sizes from $300$K up to $350$M parameters. A brief overview of the selected systems is as follows:

%\begin{itemize}
  %  \item \textbf{SSL based}:- WavLM, Hubert, and Wav2Vec2  based systems from\cite{kulkarni2024generalization} that use ECAPA TDNN as a back-end network. TCM Add\footnote{\url{https://github.com/ductuantruong/tcm_add}} \cite{truong2024temporal} incorporates a classification token to encode temporal and channel modeling information using a Conformer-based classifier. XLSR + SLS\footnote{\url{https://github.com/QiShanZhang/SLSforASVspoof-2021-DF}} used a sensitive layer selection approach to extract contextual information from XLS-R hidden states \cite{zhang2024audio}. Whisper Meso Net\footnote{\url{https://github.com/piotrkawa/deepfake-whisper-features}} \cite{kawa2023deepfake} integrates Whisper features with MesoNet using MFCC representations.
    
    % here are citations for w2v2, wavlm, whisper, ecapa, xls-r papers but might not be necessary to include \cite{wav2vec2, wavlm, radford2022robust, desplanques2020ecapa, conneau21crosslingual}.
     
 %   \item \textbf{Graph Neural Network Based}:-  For Graph based models we select AASIST\footnote{\url{https://github.com/clovaai/aasist}} \cite{jung2022aasist} processes raw waveforms directly, leveraging graph neural networks and spectro-temporal attention mechanisms. RawNet2\footnote{\url{https://github.com/eurecom-asp/rawnet2-antispoofing}} \cite{tak2021end} applies time-domain convolution operations directly to raw audio. RawGat-ST\footnote{\url{https://github.com/eurecom-asp/RawGAT-ST-antispoofing}} \cite{tak21_asvspoof} utilizes spectral and temporal sub-graphs combined with a graph pooling strategy.

%\end{itemize}

\section{Experimentation}
\label{experi}
\subsection{Experimental setup}

%We obtain readily available pre-trained weights for the $9$ selected models released in the original work. Details regarding the architecture and variants of the models used can be found in \ref{}. All the selected systems are trained on the ASVSpoof 2019 dataset except for the Whisper MesoNet system which is trained on a subset of the ASVSpoof 2021 DF dataset. 

%For the benchmarking data, we obtain the only evaluation split from all the datasets. All the audio files are sampled at 16KHz. For the WavLM, Hubert and Wav2Vec2-based systems, we pad the input audio based on the longest input in the batch. Rest of the systems operate on a fixed input length of 4s, the input signal is thus trimmed to this duration.

%%All the benchmarking experiments are conducted using the DF Arena toolkit with a batch size of 64 on a single Nvidia L40S (48GB) GPU.

% <batchsize info anf audio length info missing ??> \newline

We use readily available pre-trained weights for the open-source models, as released in their original works.  All selected systems are trained on the ASVspoof 2019 dataset, except for the Whisper MesoNet system, which is trained on a subset of the ASVspoof 2021 DF dataset. For benchmarking, we use only the evaluation split from each dataset, ensuring consistency across experiments. All audio files are sampled at $16$kHz. For WavLM, Hubert, and Wav2Vec2 \cite{kulkarni2024generalization} based systems, we pad input audio based on the longest sequence in each batch. The remaining systems operate on a fixed $4$s input length, so input signals are trimmed accordingly. All benchmarking experiments are conducted using the Speech DF Arena toolkit, with a batch size of $64$, running on Nvidia GPU variants, L40S (48GB), and A100 (40GB).

\begin{table*}[t]
\centering
\caption{EER (\%) of SOTA deepfake detection models across various datasets. Datasets described in Section \ref{dataset} and detection systems in Table 1. T in the table refers to track, and R refers to round. Proprie. refers to the proprietary systems submitted to the DF arena.} 
\resizebox{\textwidth}{!}{%
\begin{tabular}{ll|c|cccc|c|c|cc|cc|c|c|c}
\toprule
& \multicolumn{1}{c|}{\textbf{System/Dataset}}  & \multicolumn{1}{c|}{\textbf{ITW}} & \multicolumn{4}{c|}{\textbf{ASVspoof}} & %\multicolumn{1}{c}{ASV21DF} & \multicolumn{1}{c}{ASV24-E} & 
\multicolumn{1}{c|}{\textbf{FoR}} & \multicolumn{1}{c|}{\textbf{CodecFake}} & \multicolumn{2}{c|}{\textbf{ADD22}}  & \multicolumn{2}{c|}{\textbf{ADD23}} 
%& \multicolumn{1}{c}{ADD23-R1} 
& \multicolumn{1}{c|}{\textbf{DFADD}} & \multicolumn{1}{c|}{\textbf{LibriSeVoc}} & \multicolumn{1}{c}{\textbf{SONAR}} \\

& \textbf{} & {} & \textbf{2019} & \textbf{2021LA}  & \textbf{2021DF} & \textbf{2024} & \textbf{} & \textbf{} & \textbf{T1} & \textbf{T3} &\textbf{R1} & \textbf{R2} & \textbf{} & \textbf{}& \textbf{} \\
\midrule
\parbox[t]{2mm}{\multirow{3}{*}{\rotatebox[origin=c]
   {90}{\textbf{{Proprie.}}}}} 
& Whispeak & \textbf{1.26} & 0.39 & 3.58 & 3.23 & \textbf{9.92} & 1.02 & \textbf{0.86} & \textbf{11.94} & \textbf{2.31} & \textbf{2.62} & \textbf{5.01} & 0.00 & \textbf{0.08} & \textbf{0.53}  \\ 
& Syntra Detector & 3.97 & 1.48 & 14.05 & 2.02 & 15.96 & \textbf{0.53} & 1.18 & 23.58 & 2.98 & 3.65 & 9.60 & 0.00 & 1.05 & 0.65  \\ 
& Resemble Detect & 3.94 & 1.32 & 1.64 & 3.79 & 16.29 & 1.36 & 33.04 & 28.21 & 6.11 & 21.07 & 28.27 & 0.00 & 1.65 & 2.98  \\ 

\midrule
\parbox[t]{2mm}{\multirow{11}{*}{\rotatebox[origin=c]
   {90}{\textbf{Open-source}}}}
& XLSR+SLS & 7.45 & 0.23 & 2.86 & 1.91 & 18.76 & 5.07 & 33.43 & 33.95 & 15.74 & 19.37 & 21.09 & 7.54 & 1.72 & 24.72  \\ 
& TCM & 7.79 & 0.18 & 2.99 & 2.14 & 18.85 & 10.68 & 36.00 & 37.40 & 20.94 & 23.42 & 22.74 & 8.88 & 2.27 & 26.57  \\ 
& Nes2NetX & 7.75 & \textbf{0.12} & 2.17 & \textbf{1.49} & 22.05 & 6.31 & 39.34 & 34.47 & 26.55 & 21.13 & 18.44 & 11.14 & 3.04 & 31.53  \\ 
& Wav2Vec2 AASIST & 11.19 & 0.22 & \textbf{0.82} & 6.63 & 16.24 & 7.46 & 43.36 & 31.04 & 16.52 & 27.74 & 21.92 & 11.92 & 11.13 & 46.12  \\
& XLSR Mamba & 6.70 & 0.42 & 0.93 & 1.88 & 14.40 & 6.71 & 35.26 & 34.22 & 19.36 & 21.84 & 20.15 & 10.69 & 2.15 & 24.26  \\  
& Whisper Mesonet & 26.72 & 5.83 & 15.82 & 2.11 & 22.54 & 47.74 & 34.66 & 38.38 & 24.00 & 41.24 & 44.56 & 24.11 & 15.28 & 58.66  \\ 
& Wav2Vec2 ECAPA & 30.69 & 29.69 & 26.60 & 22.43 & 18.65 & 62.32 & 40.97 & 46.43 & 21.87 & 35.28 & 36.70 & 75.06 & 28.94 & 64.56  \\ 
& AASIST & 43.00 & 0.82 & 11.46 & 21.07 & 35.53 & 21.64 & 51.05 & 47.91 & 33.18 & 47.75 & 32.46 & 41.86 & 37.65 & 57.47  \\ 
& WavLM ECAPA & 34.64 & 0.76 & 6.67 & 15.94 & 25.99 & 23.36 & 46.18 & 44.16 & 39.41 & 28.99 & 31.93 & 29.53 & 32.87 & 41.94  \\ 
& RawGatST & 52.53 & 1.06 & 10.25 & 23.26 & 40.29 & 53.09 & 50.00 & 42.90 & 32.30 & 37.86 & 27.33 & 23.70 & 43.52 & 50.78  \\ 
& Rawnet2 & 49.00& 4.59 & 9.48 & 22.38 & 40.67 & 48.54 & 49.75 & 50.35 & 32.29 & 44.50 & 34.77 & 26.19 & 47.58 & 40.42  \\ 
& Hubert ECAPA & 38.65 & 1.05 & 12.55 & 13.79 & 31.39 & 33.74 & 46.22 & 47.66 & 39.08 & 49.56 & 43.95 & 34.56 & 32.21 & 40.19  \\

\bottomrule
\end{tabular}%
}
\label{tab:eer}
\vspace{-1em}
\end{table*}

\subsection{Evaluation}

We use EER as the target evaluation metric for the benchmarking. EER is the point where the false acceptance rate and false rejection rate are equal compared to a threshold. To assess the overall robustness and generalizability of a system across multiple evaluation conditions, we compute the pooled EER. The pooled EER is derived by aggregating all verification scores, both genuine and spoofed, from all datasets into a single score distribution. The decision threshold is then determined globally, providing a unified performance measure that reflects the system’s capability to generalize as a single detector across diverse conditions. We conducted the correlation analysis between different datasets and the average EER (computed across all datasets). We used various correlation metrics \cite{stepanov2024correlation, Vergara_2013, Lin1989ACC}, namely Pearson, Spearman and Kendall’s Tau, Distance Correlation Mutual Information. Each correlation metric offers a unique perspective on dataset relationships. Pearson correlation measures linear dependence, while Spearman and Kendall’s Tau focus on ranking consistency. Distance Correlation captures both linear and non-linear dependencies, making it a broader measure of association. Mutual Information quantifies shared information, and the Concordance correlation coefficient assesses both precision and accuracy in dataset alignment. Together, these metrics provide a comprehensive view of how well each dataset’s EER distribution corresponds to the overall trend. 

All the systems on the Speech DF Arena were also evaluated in terms of Accuracy and F1-score (available on the Speech DF Arena Huggingface page), providing complementary perspectives on detection reliability. 
Accuracy captures the overall proportion of correctly classified samples, making it a straightforward measure of effectiveness across balanced datasets. However, in deepfake detection tasks where the distribution of genuine and manipulated speech can be uneven, Accuracy alone may obscure critical weaknesses. The F1-score, defined as the harmonic mean of precision and recall, addresses this by emphasizing the balance between false positives and false negatives. High F1-scores therefore indicate that a system not only detects synthetic speech reliably but also avoids over-flagging authentic speech, a property crucial for real-world deployment. Together, Accuracy and F1-score complement EER by illustrating both the absolute detection capacity of a system and its robustness to class imbalance, providing a more comprehensive evaluation of detection reliability across diverse datasets.

\section{Results and analysis}
\label{results}
This section presents the performance obtained by the open-source SOTA and proprietary detection systems across the datasets. %Performance is reported in terms of EER, average EER, pooled EER, Accuracy, and F1 score.

\subsection{Performance of SOTA open-source systems}

The second row of Table \ref{tab:eer} presents the EER of the open-source systems across various datasets. We have evaluated all the systems on $14$ datasets, ensuring that models are tested on diverse deepfake generation techniques, including TTS, VC, Neural Audio Codecs, and Vocoders as described in Section \ref{dataset}. One of the key observations from Table \ref{tab:eer} is that not a single model consistently outperforms others across all datasets. XLSR+SLS emerges as the most effective open-source system among all, achieving the lowest EER except the ASVspoof series. Nes2NetX achieves the best performance in ASVspoof 2019 with an EER of $0.12\%$ and $1.49\%$ in ASVspoof2021DF, even surpassing proprietary systems, but underperforms in several datasets. RawNet-2, RawGAT-ST, Hubert and AASIST struggle significantly, with consistently high EER values, suggesting that their architectures are not able to generalize well to certain types of deepfake attacks. RawNet-2 exhibits particularly weak performance in most of the datasets, EER exceeding more than $40\%$, indicating the inability to generalize to out-of-distribution data. Similarly, RawGAT-ST achieves high EER ($52.54\%$) in the In-the-wild and ASVspoof-2021DF ($23.26\%$), making it ineffective against diverse attacks. Although AASIST achieved the best performance in ASVspoof 2021LA with an EER of $0.82\%$, it still underperforms in key datasets such as CodecFake ($51.05\%$) and ADD22-T1 ($47.91\%$).

\begin{table}[h]
    \centering
    \caption{Performance comparison of various systems in terms of number of parameters, average EER(\%) and pooled EER(\%).}
    \label{tab:average}
    \resizebox{\columnwidth}{!}{%
    \begin{tabular}{llccc}
        \toprule
        & \textbf{System} & \textbf{Num. Params. (M)} & \textbf{Average EER (\%)} & \textbf{Pooled EER (\%)} \\
        \midrule
        \parbox[t]{2mm}{\multirow{3}{*}{\rotatebox[origin=c]
   {90}{\textbf{Proprie.}}}} &
        Whispeak     & 98.90  & \textbf{3.05}  & \textbf{3.00} \\
        & Syntra.io     & 584.00  & 5.76  & 11.29 \\
        & Resemble AI     & 2112.00 & 10.69 & 12.37 \\
        \midrule
        \parbox[t]{2mm}{\multirow{11}{*}{\rotatebox[origin=c]
   {90}{\textbf{Open-source}}}} 
        & XLSR+SLS            & 340.00  & 13.84 & 15.68 \\
        & TCM                 & 319.00  & 15.77 & 16.35 \\
        & Nes2NetX            & 317.90  & 16.11 & 17.04 \\
        & Wav2Vec2 AASIST     & 317.84  & 18.02 & 19.47 \\
        & XLSR Mamba          & 319.00  & 14.21 & 20.12 \\
        & Whisper Mesonet     & 7.60    & 28.69 & 23.76 \\
        & Wav2Vec2 ECAPA      & 324.00  & 38.58 & 28.81 \\
        & AASIST              & 0.30    & 34.49 & 33.16 \\
        & WavLM ECAPA         & 102.00  & 28.74 & 33.48 \\
        & RawGatST            & 0.44    & 34.92 & 33.93 \\
        & RawNet2             & 17.60  & 35.75 & 35.66   \\
        & Hubert ECAPA        & 102.0  & 33.19  & 43.03 \\
        \bottomrule
    \end{tabular}
     }
    \label{tab:perfcomp}
    %\vspace{-0.5em}
\end{table}

Table \ref{tab:perfcomp} compares detection models in terms of average EER, pooled EER, and parameter size, highlighting trade-offs between accuracy, generalization, and efficiency. Among mid-sized transformer-based models, XLSR+SLS (340M) achieves the best balance with $13.84\%$ average EER and $15.68\%$ pooled EER, while XLSR Mamba (319M) shows similar average EER ($14.21\%$) but higher pooled EER ($20.12\%$), indicating less robust generalization. TCM (319M) and Nes2NetX (317.9M) perform slightly worse but remain reasonably consistent. Larger models like Wav2Vec2 AASIST (317.84M) do not necessarily improve performance, while lightweight models such as Whisper Mesonet (7.6M) and AASIST (0.3M) offer efficiency at the cost of high EERs. Some backbone-based models, like WavLM ECAPA and Wav2Vec2 ECAPA, show notable differences between average and pooled EER, reflecting variability across datasets. Overall, mid-sized transformer models provide the best trade-off between size, accuracy, and robustness, while extremely small or raw waveform models struggle to generalize, emphasizing the importance of balancing performance with computational constraints for deployment. Evidently, in several open-source systems, pooled EER is noticeably higher than average EER, e.g., XLSR Mamba (14.21\% → 20.12\%) and Wav2Vec2 ECAPA (38.58\% → 28.81\%). This suggests that these models may perform well on some datasets but are less consistent across pooled evaluations with a global threshold.

\begin{figure}[!t]
    \centering
    \includegraphics[width=0.9\linewidth]{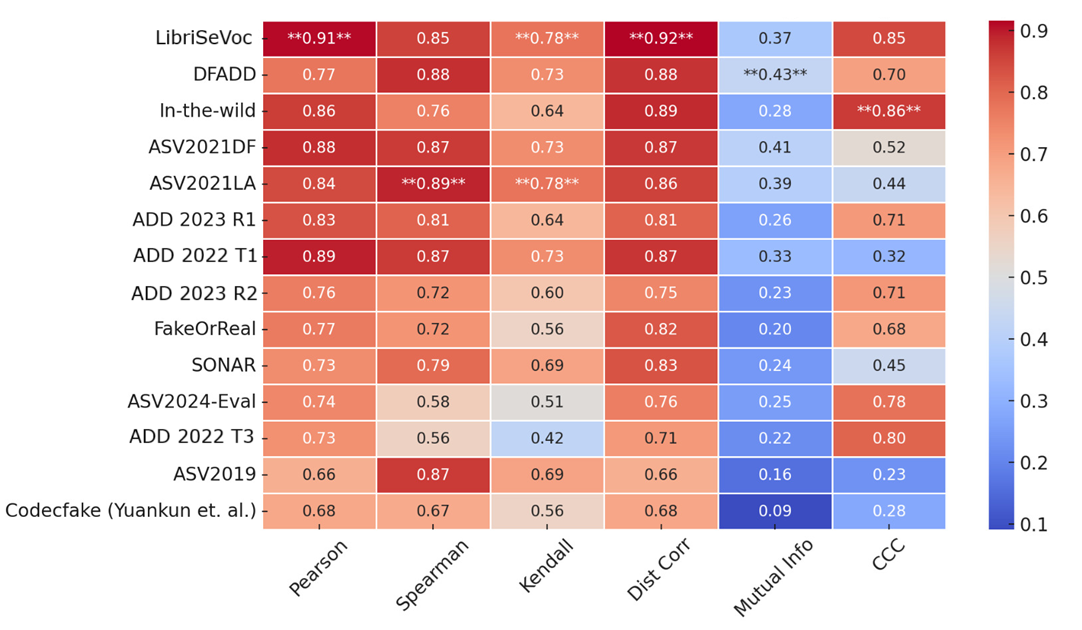}
    \caption{Correlation analysis between different datasets using heatmap; Dist.corr refers to Distance correlation, and CCC refers to Concordance correlation coefficient.}
    \label{fig:correlation}
    \vspace{-1.5em}
\end{figure}

%\vspace{-1em}
\subsection{Sensitivity study}

To study the robustness of deepfake detection models, we apply MUSAN noise perturbations from \cite{snyder2015musan} and simulated Room Impulse Response from \cite{ko2017study} to the ASVspoof 2024 dataset and evaluate the performance of the three open-source SOTA systems. Noise, speech, and music augmentations are applied with SNRs randomly sampled from $[0,15]$, $[13,20]$, and $[5,15]$ dB, respectively. The resulting EERs are presented in Table \ref{tab:augment}. Even basic data augmentation introduces significant challenges for SOTA systems, with added reverberation causing a $50\%$ relative increase in EER across all tested models. This lack of robustness highlights a critical limitation for the practical deployment of these detection systems and should be investigated further.

\subsection{Correlation analysis}

% <ADD citations for correlation measures, also add acronyms for CCC, >

We analyze the correlation of different datasets with the average EER to investigate which dataset is the best representative of the average scores, as evaluation of all the models across different datasets is a time-consuming task. The heatmap in Figure \ref{fig:correlation} provides a visual representation of these correlations. We use Pearson correlation for assessing linear dependencies and Distance correlation for assessing non-linear dependencies. We also compute mutual information to explore shared information. The heatmap suggests that LibriSeVoc and ASV2024-Eval exhibit strong Pearson, Spearman, and Distance Correlation, indicating a consistent relationship with the EER. In contrast, CodecFake and ASVspoof2019 demonstrate weaker alignment. Distance Correlation and Spearman Correlation tend to yield higher values, indicating that non-linear dependencies and ranking similarities are more prominent than strict linear correlations. Mutual information values are generally lower, indicating that some correlation
exists, the shared information between dataset-specific EERs and the average EER is limited.

\subsection{AUC Analysis}

\begin{figure}[!t]
    \centering
    \includegraphics[width=0.8\linewidth]{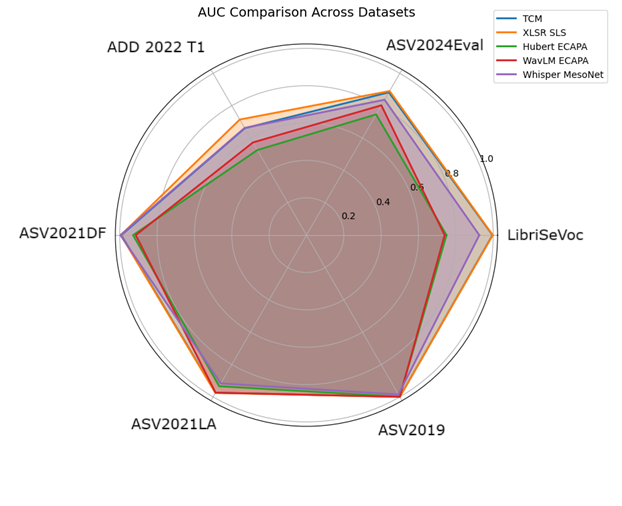}
    \caption{AUC analysis of 5 systems on ASVspoof 24, 19, 21 LA \& DF, LibriSeVoc and ADD 22 Track 1.  TCM and XLSR SLS, using the XLSR backbone and being the largest system, show the best generalization over other systems.}
    \label{fig:auc}
    \vspace{-0.9em}
\end{figure}

Figure \ref{fig:auc} presents the average on the AUC (Area under ROC curve) for the $5$ systems across $5$ highly correlated datasets. As all the systems except Whisper MesoNet are trained on ASVspoof 2019 data, we see near-perfect performance on the dataset. However, their ability to generalize to ASVspoof 2024 and ADD 2022 tasks presents a significant challenge, as indicated by the noticeable drop in performance. TCM and XLSR SLS both using the XLSR backbone and being the largest systems show the best generalizability as compared to other systems despite being trained on the same data.  

\subsection{Impact of data augmentation}

\begin{table}[!t]
 \centering
 \caption{EER (\%) of open-source SOTA systems on ASVspoof24 (ASV24) and LibriSeVoc (LibriSV) after applying various data augmentation techniques.}
 \label{tab:augment}
 \resizebox{\columnwidth}{!}{%
 
 \begin{tabular}{llccccc}
 \toprule
 & \textbf{Model} & \textbf{Original} & \textbf{Music} & \textbf{Noise} & \textbf{Speech} & \textbf{Reverberation} \\
 \midrule
 \parbox[t]{2mm}{\multirow{3}{*}{\rotatebox[origin=c]
   {90}{\textbf{ASV24}}}}
 %\multicolumn{6}{c}{\textbf{ASVSpoof}} \\
 & XLSR+SLS & \textbf{18.76} & 19.92 & 21.88 & \textbf{20.94} & \textbf{27.02} \\
 & TCM & 18.85 & \textbf{19.28} & \textbf{21.25} & 27.32 & 27.78 \\
 & WavLM-ECAPA & 24.44 & 40.67 & 38.99 & 30.74 & 35.82 \\
 \midrule
 %\multicolumn{6}{c}{\textbf{LibriSeVoc}} \\
 \parbox[t]{2mm}{\multirow{3}{*}{\rotatebox[origin=c]
   {90}{\textbf{LibriSV}}}}
 & XLSR+SLS & \textbf{1.96} & \textbf{6.89} & \textbf{9.13} & \textbf{7.01} & \textbf{21.39} \\
 & TCM & 2.34& 6.96 & 9.20 & 7.65 & 22.56 \\
 & WavLM-ECAPA & 31.88 & 51.99 & 50.77 & 39.86 & 52.29 \\ %& 39.95 & 39.01 & 30.63 & 40.18 \\
 \bottomrule
 \end{tabular}
}
\end{table}

Table \ref{tab:augment} evaluates the impact of various data augmentation techniques on the SOTA open-source detection models on ASVspoof 2024 and LibriSeVoc. On ASVspoof 2024, XLSR+SLS and TCM exhibit similar baseline performance on the original dataset with an EER of $18.76\%$ and $18.85\%$. WavLM-ECAPA performs poorly with a higher EER of $24.44\%$, indicating lower robustness. Data augmentation, such as music, noise, speech, and reverberation, degrades the performance across all the models, but the extent varies. Music and noise have a moderate impact on XLSR+SLS and TCM, with slightly higher EER. However, speech augmentation affects TCM more severely ($27.32\%$ EER), indicating sensitivity to interfering human-like sounds. Reverberation causes the most degradation across all models, with XLSR+SLS and TCM reaching around $27\%$ EER and WavLM-ECAPA degrading to $35.82\%$. Notably, WavLM-ECAPA is particularly vulnerable to music and noise augmentations, with EER spiking to $40.67\%$ and $38.99\%$, respectively. This analysis suggests that while XLSR+SLS and TCM maintain more stable performance under augmentation, WavLM-ECAPA struggles significantly, highlighting the importance of augmentation-aware training for robust deepfake detection.

For the LibriSeVoc dataset, XLSR+SLS shows a similar trend, with an EER of $1.86\%$ in the original data, which increases significantly when augmented with music ($6.89\%$), noise ($9.13\%$) and speech ($7.01\%$). Reverberation has the most substantial negative impact on this model, with the EER rising to $21.39\%$. TCM exhibits comparatively higher EERs overall, with a significant increase when exposed to reverberation ($22.56\%$). WavLM-ECAPA faces the most pronounced degradation on the LibriSeVoc dataset as well, particularly with music ($51.99\%$) and noise ($50.77\%$), demonstrating its poor adaptability to these augmentations compared to the other models. These results suggest that while larger models like WavLM-ECAPA are effective in clean conditions, they are more vulnerable to distortions introduced by data augmentation, whereas models like XLSR+SLS and TCM, although affected, tend to perform more robustly across different augmentation techniques.

\subsection{Performance of the proprietary systems}

Among the proprietary systems, the Whispeak system demonstrates consistently low EER across both curated and real-world datasets (refer to table\ref{tab:detectors}), suggesting that it integrates robust machine learning classifiers optimized for cross-domain generalization. Its strong performance in challenging conditions, such as in In-the-wild 2019 and SONAR, indicates careful use of noise-robust training strategies, making it highly reliable in practical scenarios. Whispeak with $98.9$M parameters achieved the best performance in terms of average EER ($3.05\%$) and pooled EER (3\%). This indicates that it achieves strong accuracy without requiring an extremely large model. The Syntra.io system similarly achieves strong detection accuracy, particularly in the Fake or Real dataset with the lowest EER of $0.53\%$ (refer to table\ref{tab:detectors}), while maintaining sensitivity to synthetic artefacts in controlled datasets. Its ability to sustain low EERs across diverse conditions suggests a design that balances fine-grained feature representation with effective regularization, enabling better robustness to codec distortions and unseen manipulations compared to many open-source baselines.

The Resemble AI system performs competitively on curated datasets such as ASVspoof 2021LA and FoR 2021DF but exhibits larger degradation on multi-round and partially manipulated datasets like ADD23. This performance implies that the system excels at identifying global synthetic artifacts but is less optimized for localized or subtle manipulations, reflecting a gap in its generalization strategy despite its strong capabilities in clean synthetic environments. Both Syntra.io ($5.76\%$ average EER, $11.29\%$ pooled EER) and Resemble AI ($10.69\%$ average EER and $12.37\%$ pooled EER) are larger models (584M and 2112M parameters, respectively) but do not surpass Whispeak, suggesting that model size alone does not guarantee superior performance. The higher pooled EER, especially for Syntra.io, indicates limited robustness or generalization. These models may be overfitting to specific conditions or datasets and are less consistent when exposed to broader data variability. The pooled EER also suggests that larger size does not ensure both accuracy and generalization; model architecture, training strategy, and data diversity are crucial for achieving stable performance.

%\vspace{-1em}
\section{Conclusion and future work}
\label{conclude}

\begin{figure}
    \centering
    \includegraphics[width=0.7\linewidth]{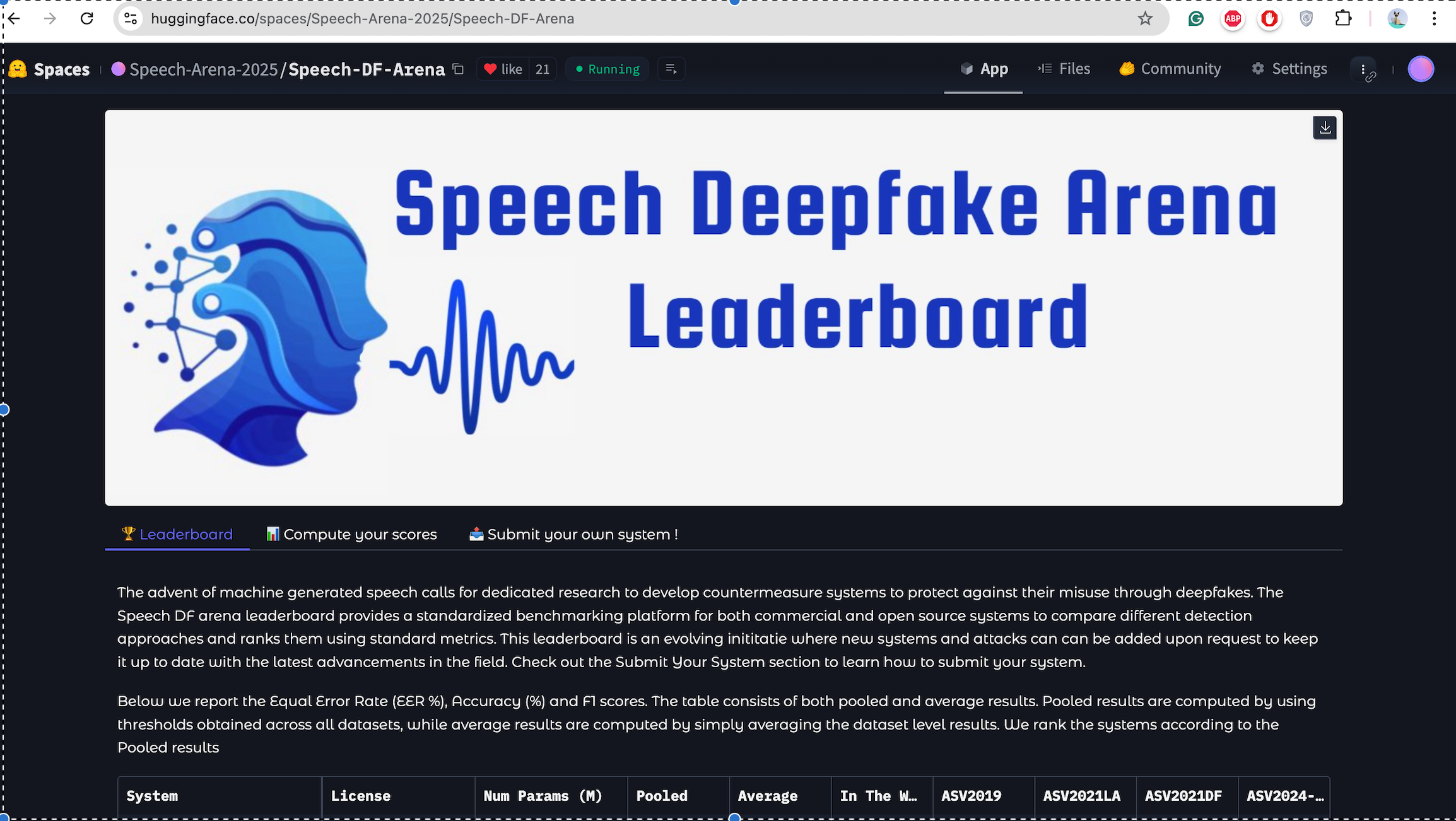}
    \caption{Interface of Speech DF Arena hosted on Huggingface. }
    \label{fig:placeholder}
    \vspace{-1.5em}
\end{figure}
%We developed Speech DF Arena, an unified and comprehensive platform by benchmarking different deepfake detection models to ensure consistent, trustworthy, and transparent evaluation across different datasets. The publicly available leaderboard on HugginFace\footnote{\url{https://huggingface.co/spaces/Speech-Arena-2025/Speech-DF-Arena}} is introduced for model comparison and selection accessible to everyone. We hope that the Speech DF Arena tool enables researchers and developers to evaluate detection systems under identical conditions, ensuring fair evaluations. Most of the detection systems perform well on particular datasets but struggle when exposed to unseen out-of-distribution deepfake attacks. By testing across multiple datasets, benchmarking helps assess the generalizability and robustness of these systems in real-world scenarios. Moreover, benchmarking fosters collaboration by providing a transparent and competitive environment, encouraging innovation and improvements in audio deepfake detection.

We developed Speech DF Arena, a unified and comprehensive platform for benchmarking audio deepfake detection models, ensuring consistent, trustworthy, and transparent evaluation across diverse datasets. To facilitate model comparison and selection, we introduced a publicly available leaderboard on HuggingFace, making it accessible to everyone. Our goal is to provide researchers and developers with a standardized evaluation framework, enabling fair and reproducible assessments of detection systems under identical conditions. While many models perform well on specific datasets, they often struggle against unseen, out-of-distribution attacks. By testing across multiple datasets, the Speech DF Arena helps assess the generalizability and robustness of these models in real-world scenarios. We hope that our benchmarking platform fosters collaboration by creating a transparent and competitive environment, driving innovation and advancements in audio deepfake detection.

%- Brining transperncy and more openness to sota systems \\
%\vspace{.1em}
\noindent\textbf{Future work}
In Phase I of the leaderboard, we benchmarked $15$ systems, including proprietary systems across $14$ audio deepfake datasets. Future expansions will include more recent datasets, covering a broader range of languages beyond English and Chinese. To enhance reliability, we plan to release evaluation sets without labels to mitigate the risk of data leakage in the evaluation pipeline. We will also allow users to test various deepfake detection systems as a live demo on HuggingFace. By integrating these enhancements into the validation process, the Speech DF Arena seeks to improve the robustness and trustworthiness of evaluations, ultimately enabling more reliable deployment of speech models in real-world applications. %Future iterations will extend benchmarking to partial spoofing scenarios, further strengthening the evaluation framework.
\section{Acknowledgement}
This work was partially supported by the Swiss National Science Foundation project Pathological Speech Synthesis (PaSS)'' (grant agreement no. 219726), and the Innosuisse flagship project Inclusive Information and Communication Technologies (IICT) (grant agreement no. PFFS-21-47). This work was partially supported by the Estonian Centre of Excellence in Artificial Intelligence (EXAI).
%\pagebreak
\bibliographystyle{IEEEtran} % or your style
\bibliography{template} % basenames only

\end{document}